\documentclass[%
 aip,
 jmp,%
 amsmath,amssymb,
 reprint,%
]{revtex4-2}

\usepackage{graphicx}
\usepackage{dcolumn}
\usepackage{bm}
\usepackage{url} 

\preprint{ }
 \usepackage[version=4]{mhchem}
\newcommand{\blue}{\textcolor{blue}}

\usepackage[normalem]{ulem}
\usepackage{graphicx}
\usepackage{dcolumn,xcolor}
\usepackage{bm}
\usepackage{url} 
\begin{document}

\preprint{}

\title[Diamond nano-micro-structures]{Diamond membranes: platform for photonic and opto-mechanical applications}
\author{
Hsin-Hui Huang$^{1,2,3}$, Gediminas Seniutinas$^{4*}$, Haoran Mu$^{1,3}$, Nguyen Hoai An Le$^1$, Eulalia Puig Vilardell$^{5,6}$, Vijayakumar Anand$^1$, Jitraporn Vongsvivut$^7$, Tomas Katkus$^1$, Meguya Ryu$^{8,*}$, Junko Morikawa$^{8,9,10}$, Saulius Juodkazis$^{1,5,9}$}%

\affiliation{Optical Sciences Centre, Swinburne University of Technology, Hawthorn, Victoria 3122, Australia}
\affiliation{Australian Research Council (ARC) Industrial Transformation Training Centre in Surface Engineering for Advanced Materials (SEAM), Swinburne University of Technology, Hawthorn, VIC, 3122, Australia}
\affiliation{Melbourne Center for Nanofabrication (MCN), 151 Wellington Road, Clayton, Vic 3168, Australia}
\affiliation{Paul Scherrer Institut, Villigen PSI 5232, Switzerland
}
\affiliation{Laser Research Center, Physics Faculty, Vilnius University, Saul\.{e}tekio Ave. 10, 10223 Vilnius, Lithuania}
\affiliation{Institute of Physics, University of Tartu, W. Ostwaldi 1, 50411 Tartu, Estonia}
\affiliation{Infrared Microspectroscopy (IRM) Beamline, ANSTO‒Australian Synchrotron, 800 Blackburn Road, Clayton, Victoria 3168, Australia}
\affiliation{~School of Materials and Chemical Technology, Institute of Science Tokyo, 2-12-1, Ookayama, Meguro-ku, Tokyo 152-8550, Japan}
\affiliation{~World Research Hub (WRH), School of Materials and Chemical Technology, Institute of Science Tokyo, 2-12-1, Ookayama, Meguro-ku, Tokyo 152-8550, Japan}
\affiliation{~Research Center for Autonomous Systems Materialogy (ASMat), Institute of Innovative Research, Institute of Science Tokyo,Yokohama 226-8501, Japan}



\date{\today}

\begin{abstract}
Diamond $1-10~\mu$m thick membranes are platform for photonic, quantum and optomechanic devices with applications across UV-IR spectral ranges. IR characterization of diamond gratings in reflection and transmission showed a change of the IR absorbance dichroism between positive and negative when the grating period was 1-2 wavelengths (free space) including inside the region of the intrinsic diamond absorbance. 
Femtosecond laser cutting of micrometers-wide and mm-long structures are demonstrated by steps of carbonization $>0.4$~J/cm$^2/$pulse (1030~nm/200~fs) and oxidation of diamond membranes. Light intensity distribution inside form-birefringent diamond structure was modeled for a scaled-down structure and wavelength to reveal characteristic interference patterns for different polarizations.   
\end{abstract}

\maketitle
\tableofcontents
\begin{quotation}
\end{quotation}

\section{\label{intro}Introduction}

Advances in nanophotonics and optomechanics have highlighted the dual role of light, both as a tool for manipulating micro- and nanoscale objects through its linear and angular momentum, and as a degree of freedom that can be precisely engineered using form-birefringent structures.
Photon or radiation pressure $I/c$, a linear momentum transfer of absorbed light (intensity $I$ and velocity of light $c$), laser tweezing via the gradient of intensity of the tightly focused light, allow manipulation of micro objects. For example, birefringent liquid crystal droplets can be trapped, moved, aligned~\cite{99pps665,Kishan}. Spin and orbital angular momenta imparts torsional forces on birefringent micro-nano objects~\cite{99apl3627} with increased complexity of internal structural/mechanical changes~\cite{05s656,09mclc143}, chiral sorting~\cite{Georg}, volume-phase transitions in gels~\cite{00n178}, optical switching~\cite{99apl3627}, and other related phenomena. Even nonlinear two-photon absorption can be induced using cw-laser tweezers in dye-doped liquid crystals~\cite{08mclc310}. Further demonstration of opto-mechanical applications include sensing of magnetic fields using a laser-trapped assembly of nano-diamonds with nitrogen vacancy NV$^-$ centers~\cite{Peter}. Cooling of a micro-particle trapped in a circularly-polarized tweezers in vacuum is example of opto-mechanical effect~\cite{cool}. On the other hand, fabrication of birefringent patterns based on a grating structure allows to engineer form birefringence $\Delta n = n_e-n_o < 0$ for manipulation of spin and orbital states of light, as well as to impart mechanical action on a micro-nano structures/objects~\cite{Krauss}; here $n_e$ denotes the extraordinary refractive index (E-field is along the optical axis) and $n_o$ is the ordinary refractive index. 

The two quests to manipulate light's spin-orbital state as well as to apply mechanical action by light carrying an angular momentum  requires to create a toolbox to make gratings and structures on micrometers-thin membranes, which can be incorporated into micro-chips and photonic devices. This is already a 3D challenge for nanolithography. The geometry of nano- and micro-structures defines their mechanical response through their moment of inertia and resonance frequency, and requires demanding fabrication approaches, especially for micrometers-thin membranes. In particular, long and thin structures are very sensitive to mechanical forces, which makes them excellent sensitive tools for opto-mechanical measurements, but also challenging for microfabrication, e.g., by laser machining using ablation. As an example, a suspended bridge structure has 
the area moment of inertia for a rectangular section $I_M=H^3W/12$ and under load $Q$ (force per unit length) deflects by $\delta_{max} = \frac{QL^4}{764EI_M}$, where $L,H,W$ are length, thickness, width of the beam and $E$ its Young modulus. 
There is a need for new methods to structure diamond membranes for photonic applications over a wide spectral range from X-ray to far-IR and for quantum emitters/sensors, spin-orbital converters (q-plates), and opto-mechanical structures. 

Fabrication of intricate sub-wavelength patterns on thinner membranes with thickness $\leq 10~\mu$m by usual EBL definition and plasma etching becomes challenging due to usually non-flat substrates which have inherent internal stresses built in during the growth or transfer and fixation onto sample mounts, e.g., Si substrates with central holes. Crystalline substrates thinner than $\sim 20~\mu$m are usually not flat, e.g., Si solar cells with efficient light trapping using  photonic crystal (PhC) patterns on $\sim 10~\mu$m wafers a promising for the above Lambertial limit performance~\cite{SJ18,SJ19,SJ20}, 
however their realization by standard lithography approaches is challenging. Hence, a light trapping PhC pattern on a 10~$\mu$m-thick Si solar cell was prototyped on Si-on-Insulator (SoI) substrate by EBL and etching, first, followed by the final removal of the Si back-substrate by wet or dry etching~\cite{SJ,21oea210086}. Recently, fs-laser patterning of sub-wavelength holes in alumina hard-etch-mask was demonstrated for PhC light trapping Si solar cells over areas with 2~cm side length~\cite{25aome01781}. The fs-laser direct write over cm-scale length with sub-1~$\mu$m precision and resolution has become a feasible method~\cite{ZHANG,ma17030557}. It can be applied to diamond (usually cubic), recently synthesized in a harder hexagonal lonsdaleite phase previously observed only in materials with meteorite impact~\cite{Wenge1,noWenge}. 

Here, we focus on two approaches to structure diamond membranes by lithography and laser patterning. First, we characterize the response of diamond gratings fabricated on a $\sim 10~\mu$m membrane at IR spectral range. Diamond gratings with azimuthally changing orientation are optical elements which control spin-orbital coupling and must be fabricated on thin membranes for opto-mechanical applications. Secondly, we characterize the structuring with fs-laser of diamond membranes of only $\sim 1~\mu$m thickness. The fabrication and characterization toolbox for such elements can be extended through high-precision machining using fs-lasers. A structure-controlled change between spectral regions with positive and negative dichroism of form-birefringent gratings is revealed.

\section{Samples and Methods}
\subsection{Diamond micro-membranes}

\begin{figure*}[tb]
\centering\includegraphics[width=1\textwidth]{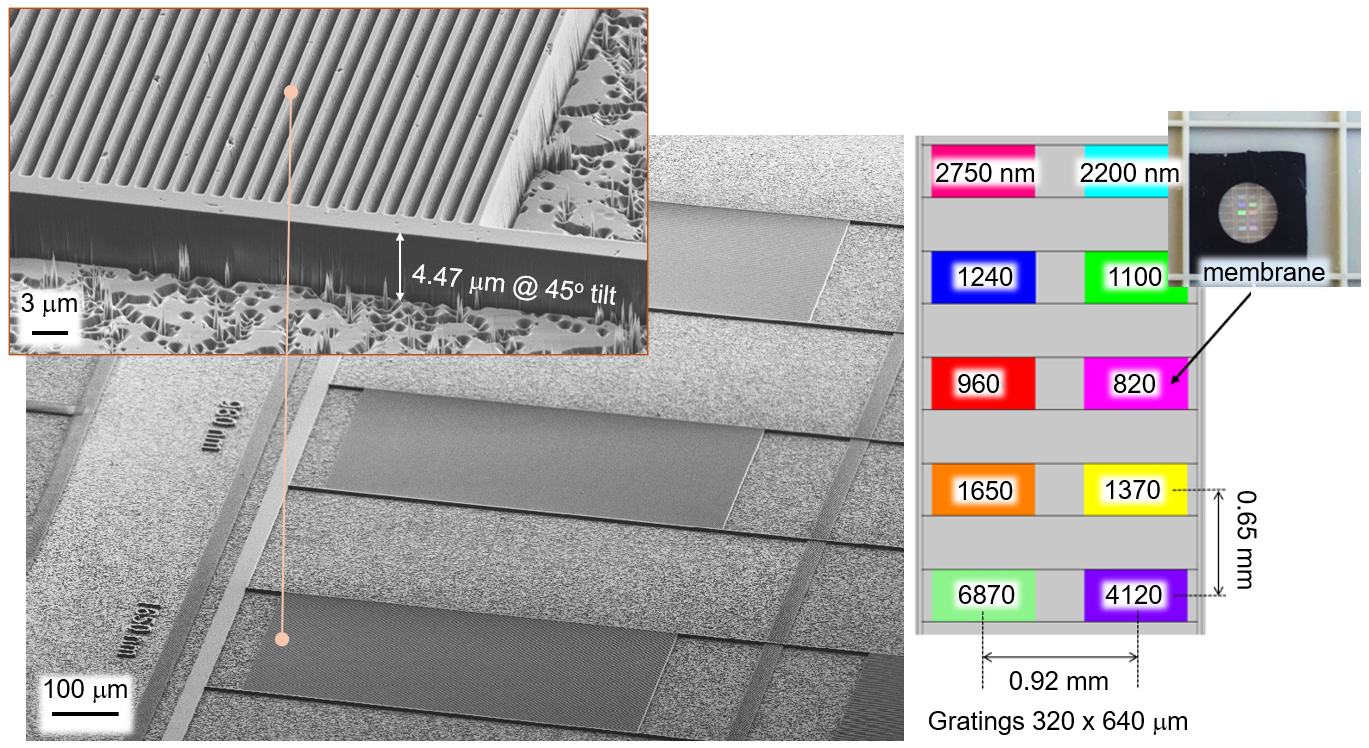}
\caption{\label{f-x} X-ray gratings in diamond on a $\sim 9.5~\mu$m free-standing membrane; diameter of the membrane circular window 5~mm. Gratings were defined by e-beam lithography (EBL) and plasma etching. The 6.32~$\mu$m high gratings of duty cycle 0.5 with different periods from $\Lambda = 820$~nm to 6.87~$\mu$m. }
\end{figure*}

Thick $\sim 9.5~\mu$m (Fig.~\ref{f-x}) and thin $\sim 1~\mu$m CVD diamond membranes from Diamond Materials GmbH were used in this study. The etch masks of HSQ-resist for gratings were defined by e-beam lithography (EBL) and made by reactive plasma etching (RIE) with details given elsewhere~\cite{Gediminas}. For thin 1-$\mu$m diamond membranes, fs-laser micro-machining by graphitization-oxidation and by ablation were made directly on as-received diamond membranes. The bandgap of diamond is $E_g=5.45$~eV (wavelength $1.24~\mathrm{[eV]}/E_g = 228$~nm). Diamond membranes were without dopants and purely dielectric. This was also excluding presence of a prominent Reststrahlen IR band where the permittivity $\varepsilon=\varepsilon'+\varepsilon"$ has spectral region where surface wave can be launched at the air-diamond interface, the surface phonon polariton (SPhP) with $\varepsilon'<0$.    

\subsection{Femtosecond laser machining}

Diamond $1-\mu$m-thick membranes were laser cut using Mitutoyo $5^\times$, $NA = 0.14$ NIR objective lens for stress relief segments and $20^\circ$ $NA = 0.45$ NPlan Apo Olympus objective lens for the definition of beams/bridges.

Laser (Pharos, Light Conversion Ltd.) power 2.5~W and 200~kHz frequency were kept constant, wavelength was 1030~nm, pulse duration 230~fs. Process variables were attenuator transmittance $T\%$ to define pulse energy, number of pulse per site $n$, pulse density per millimeter $d_p$~mm$^{-1}$, and write speed $v_s$~[mm/s].

For stress relief cuts: progressively nibble away material on two sides symmetrically having set the remaining membrane support width to 2.5~mm, then down to 0.5~mm in 0.5~mm decrements. This allowed internal compressive stresses in the membrane to relax partially and reshape the membrane from a roundish bulge to a unidirectional bend with a comparatively much flatter crest where the platform cut was then positioned.
One cut pass and one redundancy pass were used. Pulse energy 
$E_p = 1387$~nJ corresponding to the $F_p = 2.2$~J/cm$^2$ fluence, which is slightly below single pulse ablation of diamond at 3~J/cm$^2$. The effective laser frequency defined by number of pulses per site (selected by laser pulse picker) $n = 25$, density of pulses $d_p = 300$~mm$^{-1}$, and scan speed $v_s = 2$~mm/s was $f_{ef}= n d_p v_s = 15$~kHz.

For the platform with bridges/beams cuts: 300~$\mu$m diameter central platform and 1~mm beams on two sides L and R. Bridges were cut and gradually thinned down from 40~$\mu$m to 10~$\mu$m in small steps. Periodically measurements of sag (bridge end and platform Z difference) and tilt (L-R bridge end Z difference) were taken and adjusted for in fab code. The 10~$\mu$m bridge width was the point where the platform could start shifting or flop its bulge up/down at the used laser cut conditions.

The conditions for the cut of the final structure were chosen at tighter focusing to reduce graphitisation at the rim of the cut. 
The $NA = 0.45$ was used with two passes 
at pulse energy of 185~nJ, the average fluence of 3~J/cm$^2$ and effective frequency of $f_{ef} = 200$~kHz. Extra trimming/narrowing of the bridge was carried out by additional two laser runs on the same path but at higher pulse energies of 229~nJ and 451~nJ.
\begin{figure*}[tb]
\centering\includegraphics[width=1\textwidth]{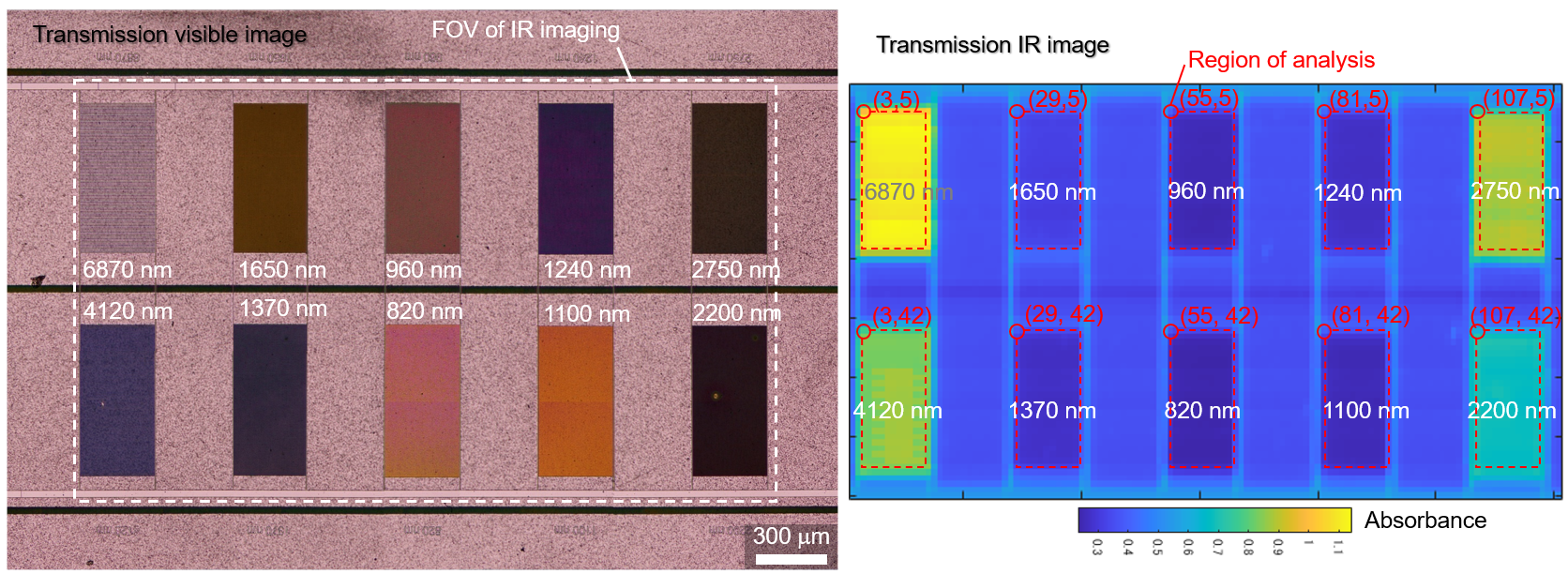}
\caption{\label{f-OIR} Optical and IR transmission images of the gratings (Fig.~\ref{f-x}). Absorbance $A = -\lg T$, where $T$ is the transmittance. The IR absorbance map is measured over the entire spectral range $4000 - 1000$~cm$^{-1}$; measured with $NA = 0.6$ objective lens in the imaging mode with a $25~\mu$m raster step (Spotlight, PerkinElmer). Regions of analysis are marked with dashed-box markers over the entire grating areas.}
\end{figure*}

\subsection{Numerical simulations}

Finite-difference time-domain (FDTD) calculations were performed using the Lumerical-Ansys software package to model the optical field distribution inside the form-birefringent diamond grating. Diamond was treated as a lossless dielectric with refractive index of $n = 2.4$.

To reduce computational cost while preserving electromagnetic similarity, a $10^\times$ geometrical down-scaling was applied according to Maxwell’s scaling invariance. The simulated grating had a period $\Lambda$ = 412~nm and duty cycle of 0.5, supported on a diamond membrane. The computational domain was periodic in the lateral $(x,y)$ directions and terminated with perfectly matched layers (PML) along the propagation (z) direction.

A plane-wave source was used with normal incidence ($\theta = 0^o$). Broadband simulations were carried out over the wavelength range $\lambda = 300\sim 1000$ nm. Linear polarization azimuths of $0^\circ$, $45^\circ$, $90^\circ$, and $135^\circ$ were simulated. The incident field amplitude was normalised such that the incident intensity satisfied $E_0^2$=1.

Reflection and transmission were recorded using 2D frequency-domain power monitors normal to the propagation direction. The reflection monitor was positioned 200~nm away from the injection plane to avoid overlap with the source field. The transmission monitor was placed 630~nm away from the exit interface to ensure full formation of the transmitted field. For front- and back-illumination cases, monitor distances from the respective grating interfaces were kept identical to ensure symmetric comparison.

Field monitors were used to record the total electric field intensity 
$E^2$ as well as the normal field component $E_x^2$, corresponding to the component perpendicular to the grating sidewalls. These distributions were analysed to identify polarization-dependent energy localisation within the diamond and air regions.

A non-uniform mesh was employed with local mesh refinement in the grating region (maximum mesh step of 4~nm) to accurately resolve field enhancement at diamond–air interfaces. The simulation time was set to 2000~fs, and convergence was ensured using an auto-shutoff threshold of $10^{-6}$.

\section{Results}

\subsection{Thick $\sim 10~\mu$m diamond membranes}

The X-ray gratings~\cite{Gediminas} were characterised for transmittance $T$ (absorbance $A=-\lg T$) and reflectance $R$ at mid-IR spectral range of $4000 - 680$~cm$^{-1}$ wavenumbers or 2.5 - 14.7$\mu$m wavelengths, respectively (Fig.~\ref{f-x}). The gratings were made on a free-standing $\sim 9.5~\mu$m-thick membrane of poly-crystalline diamond and had height of $\sim 6.32~\mu$m at periods from $\Lambda = 0.82~\mu$m to 6.87~$\mu$m, corresponding to the aspect ratio $AR =\frac{H}{\Lambda/2} \simeq 15 - 1.8$. The gratings were defined by e-beam lithography (EBL) and were plasma etched. Images of samples at visible and IR spectral ranges in transmission are shown in Fig.~\ref{f-OIR}. 

\begin{figure*}[tb]
\centering\includegraphics[width=1\textwidth]{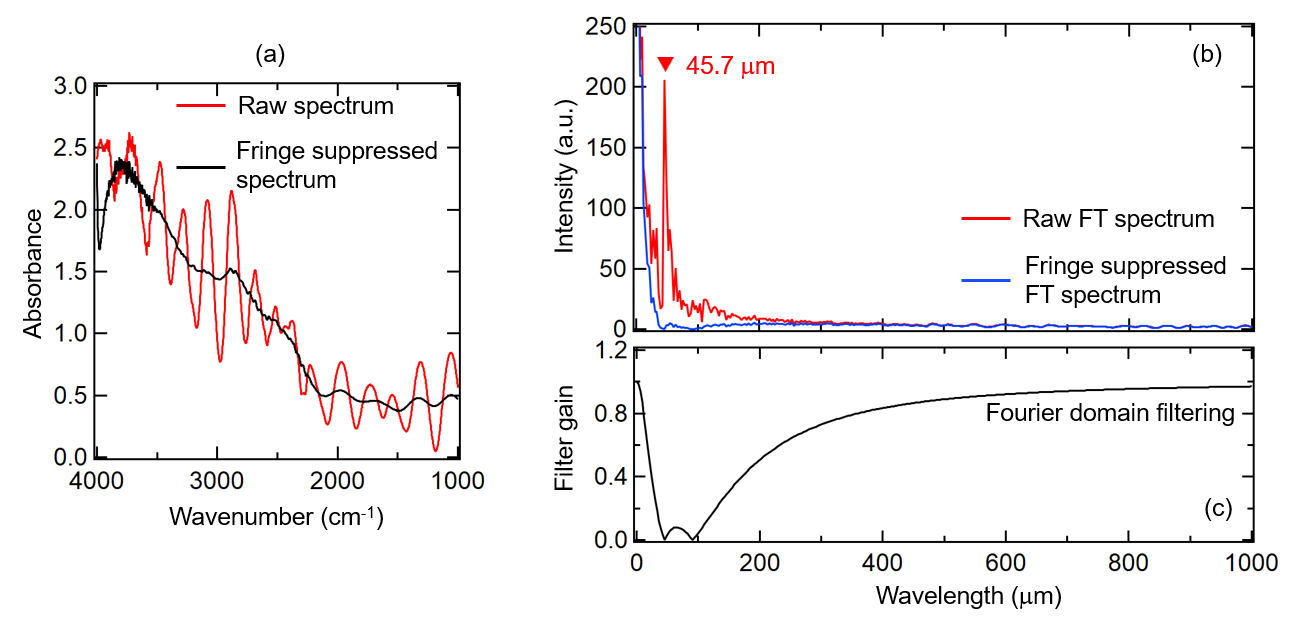}
\caption{\label{f-filt} Fringe suppression by Fourier filter. (a) IR absorbance of 2.2~$\mu$m period grating as measured and filtered. (b) The Fourier spectrum of the measured absorbance with the main free spectral range (FSR) periodicity at 45.7~$\mu$m, which corresponds to the Fabry-P\'{e}rot etalon thickness of $d = \frac{FSR}{2n} = 9.6~\mu$m for the refractive index $n = 2.38$. (c) Spectral gain profile of the filter applied for all experimentally measured absorbance $A$ spectra. }
\end{figure*}

\begin{figure*}[tb]
\centering\includegraphics[width=1\textwidth]{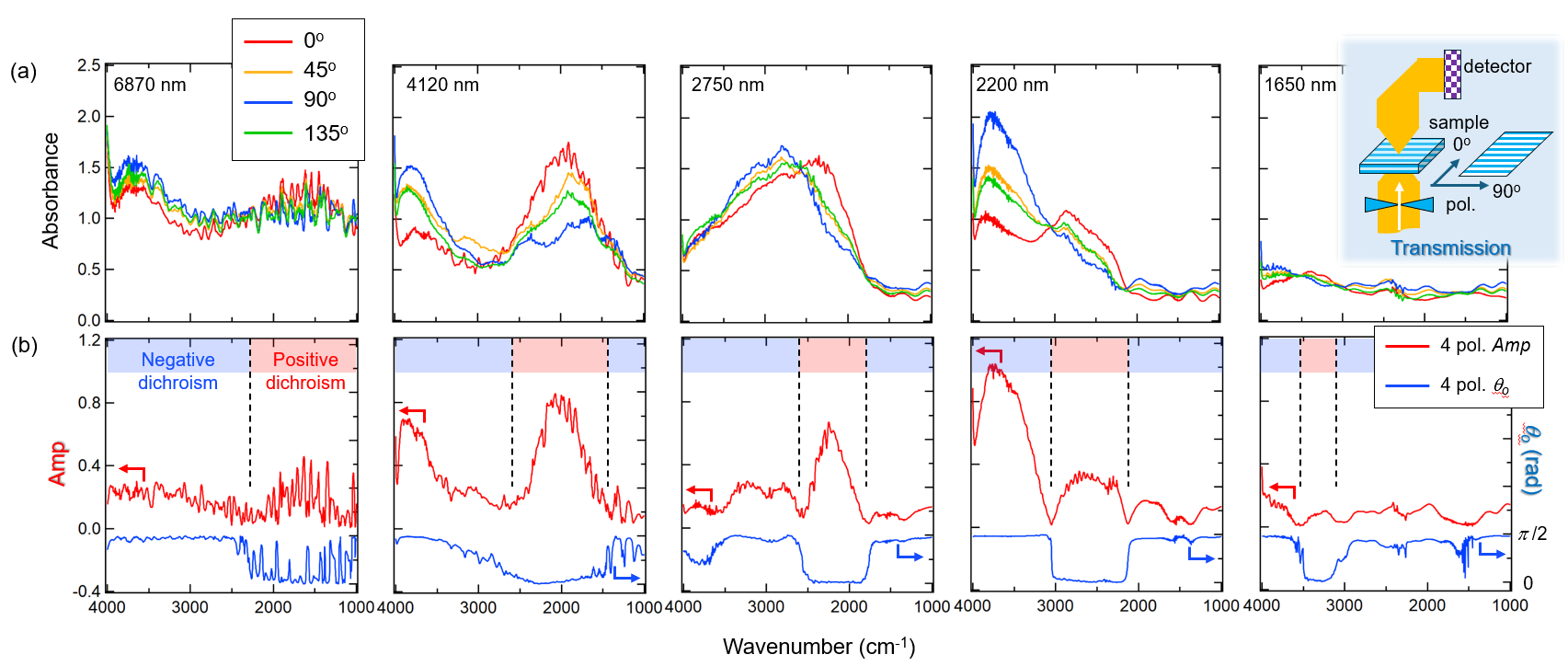}
\caption{\label{f-fin} (a) Absorbance spectra of gratings with five largest periods after fringe removal measured in transmission at azimuths $\theta = 0, \pi/4, \pi/2, 3\pi/4$ (see inset in (a)). (b) The amplitude $Amp$ and orientation angle $\theta_o$ from the fit by $Amp\times\cos(2\theta - 2\theta_o)+O$ from data at 4-pol. azimuths, where $\theta$ is the azimuthal angle, $Amp$ is the amplitude and $O$ is the offset (not plotted here). The 4-pol. fit was carried out at each pixel $25\times 25~\mu$m$^2$ level. } 
\end{figure*}

When period of the grating is comparable with the IR wavelength of incident light $\Lambda\sim\lambda$, the structure imposed changes of reflectance and transmittance are modulated/altered. The intrinsic absorption of diamond stretch over 2600-1700~cm$^{-1}$ band (3.85-5.88~$\mu$m). First, a Fourier filter was applied for experimentally measured absorbance spectrum (Fig.~\ref{f-filt}). The very same filter was applied for all absorbance spectra. Only five gratings with the most strong absorbance at IR (the largest periods) were analysed (see Fig.~\ref{f-OIR}). Figure~\ref{f-fin}(a) shows filtered absorbance spectra of five gratings at four polarization azimuths. At some spectral region, absorbance at $90^\circ$ was larger than at $0^\circ$ as well as the opposite tendency was present. The dichroic ratio $R_d = \frac{A_\parallel}{A_\perp}$ is called positive when $R_d > 1$ and negative when $R_d < 1$ (see Sec.~\ref{disco}). From the 4-pol. fits, the amplitude $Amp$ and orientation azimuth $\theta_0$ were calculated (Fig.~\ref{f-fin}(b)). The $\theta_0$ shows jumps of phase by $\pi/2$ at the locations of the sign change of the dichroism. 
For example, the iso-polarisation point where absorbance for all four polarisations is the same was at 3000~cm$^{-1}$ for the $\Lambda = 2.2~\mu$m period grating (Fig.~\ref{f-fin}). This corresponds to the wavelength of $3.33~\mu$m, which is inside diamond with the refractive index of $n = 2.35$. The isotropic absorbance for the linear polarised incident light would correspond to the quarter-wavelength conditions induced between o- and e-beams due to the form-birefringence $\Delta n$ in the grating of thickness $d = 6.32~\mu$m, i.e.  $\frac{\lambda/n}{4} = \Delta n d$. One would find $\Delta n = 0.056$, which match the expected value for the grating with duty cycle $\sim 0.5$ (the largest birefringence). The phase conditions for a quarter wavelength are repeated over the wavelength (or $2\pi$ in phase): $\frac{(2j-1)[\lambda/n]}{4}$ where $j$ is an integer $j = 1,2,..$. For $j=2$ and wavelength $\lambda = 3.33~\mu$m, it is $\frac{3\lambda}{4} = 2.498~\mu$m (or 4003~cm$^{-1}$) where spectral iso-crossing is expected (Fig.~\ref{f-fin}). This analysis is applied to a spectral range out of the intrinsic diamond absorbance 2600-1800~cm$^{-1}$ band; at the absorbance resonance the phase of an oscillator undergoes a $\pi/2$ change. 


As measured spectra of absorbance and reflectance without fringe removal are shown in Supplement. Figure~\ref{f-6780} shows $A$ and $R$ spectra from the very same area of $100\times 100~\mu$m$^2$ at four polarizations. The reduced absorbance $A$ was observed at $\Lambda\sim\lambda$ for 6.87~$\mu$m (Fig.~\ref{f-6780}) and 4.12~$\mu$m (Fig.~\ref{f-4120}) conditions. In the case of those two periods, the intrinsic diamond band is not discernible due to reduced absorbance (reduced $A$ can also be affected by diffraction losses). By color-tracing of the spectra measured at different azimuth of incident polarization, a clear inversion of phase (an intensity reversal between $0^\circ$ and $90^\circ$ spectra) was observed in Fig.~\ref{f-4120}(a) at the intrinsic diamond absorption region. For the smaller period grating of 0.82~$\mu$m, the $A,R$ spectra are governed by interference and clear free spectral range $FSR\simeq 236$~cm$^{-1}$ is observed (Fig.~\ref{f-820}). The effective refractive index $n$ can be determined for the known thickness $d$: $FSR = \frac{1}{2nd}$~cm$^{-1}$; for $d = 9~\mu$m, $n \simeq 2.35$.

\begin{figure*}[tb]
\centering\includegraphics[width=1\textwidth]{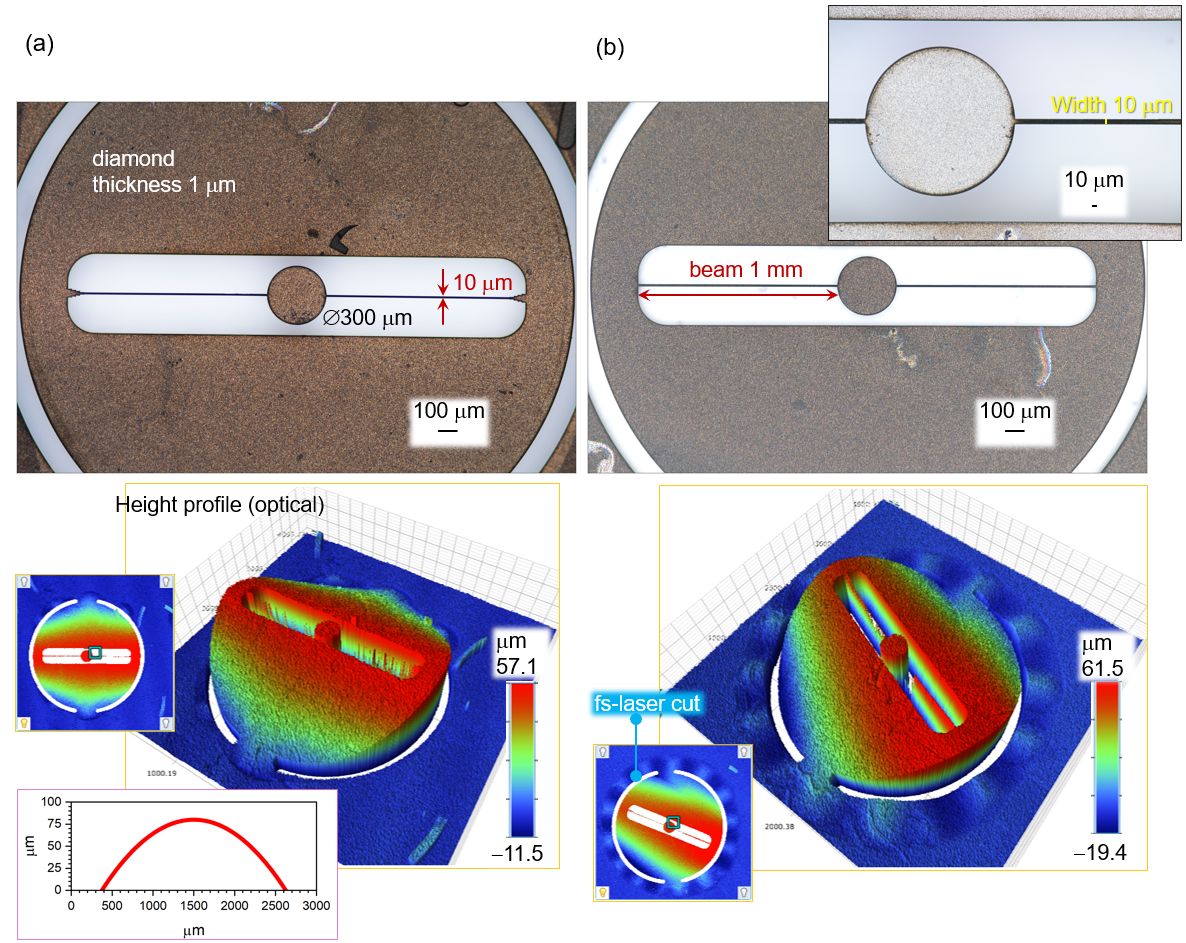}
\caption{\label{f-cut} Laser cut-out of opto-mechanical structure in a 1-$\mu$m-thick diamond membrane (free-standing over $\sim 3$~mm diameter hole). Two structures shown in (a) and (b) have different flattening pattern depending on laser semi-circular incisions to release stress; slight different in the beam anchoring sites between (a) and (b). Laser used: 230~fs/1030~nm at 200~kHz (Pharos in WOP fs-fab station). Focusing of the suspended beam-disk structure: objective lens Olympus $20^\times$ magnification $NA = 0.45$, focal diameter $2r = 1.22\lambda/NA = 2.8~\mu$m; the pulse energy 185~nJ, the average fluence of 3~J/cm$^2$ (average irradiance/intensity 13~TW/cm$^2$ per pulse), two passes. For the left structure, two extra passes for beam narrowing were carried out at 229 and 451~nJ (see Fig.~\ref{f-edge}).   
}
\end{figure*}

Interestingly, $R>1$ was observed when reflectance was normalised to that of gold mirror (on the sample stage of Bruker microscope). This can be caused by interference and sub-wavelength nature of the structure allowing for constructive field addition in intensity calculation $I\equiv (E_1+E_2+...)^2$. Strong interference is present in $A,R,T$ spectra due to the Fabry-P\'{e}rot slab multi-reflections. This hampers visualisation of dichroic ratio the dichroic ratio $R_d = \frac{A_\parallel}{A_\perp}$ change from positive to negative, however, we used experimentally measured data without Fourier filtering to trace intricate changes of absorption and reflection (see Sec.~\ref{disco} for discussion of $R_d$). With Fourier filtering, the sign change of dichroic ratio was present at the same spectral regions, however, there were oscillatory artifacts (not shown here for brevity).  

The absorbance $A$ can be estimated from the differential reflectance of the diamond grating $R_G$ vs. diamond membrane $R_D$ as $\frac{\Delta R}{ R_D} = \frac{R_G-R_D}{R_D} = \frac{4A}{n_D^2 - 1}$,~\cite{16semsc221} 
where $n_D = 2.35$ is the refractive index of diamond. This method is used to calculate absorption of a thin micro-film on a thicker substrate, e.g., resist on a cover glass. Figure~\ref{f-ref} shows such presentation for the diamond grating sample. It represents the grating-only contribution to the absorbance. The interference fringes is the main feature, however, there is a clear orientation change between the positive and negative linear dichroism at the diamond absorption region $\frac{A_{90}}{A_0} > \frac{A_{0}}{A_{90}}$.


\begin{figure*}[tb]
\centering\includegraphics[width=1\textwidth]{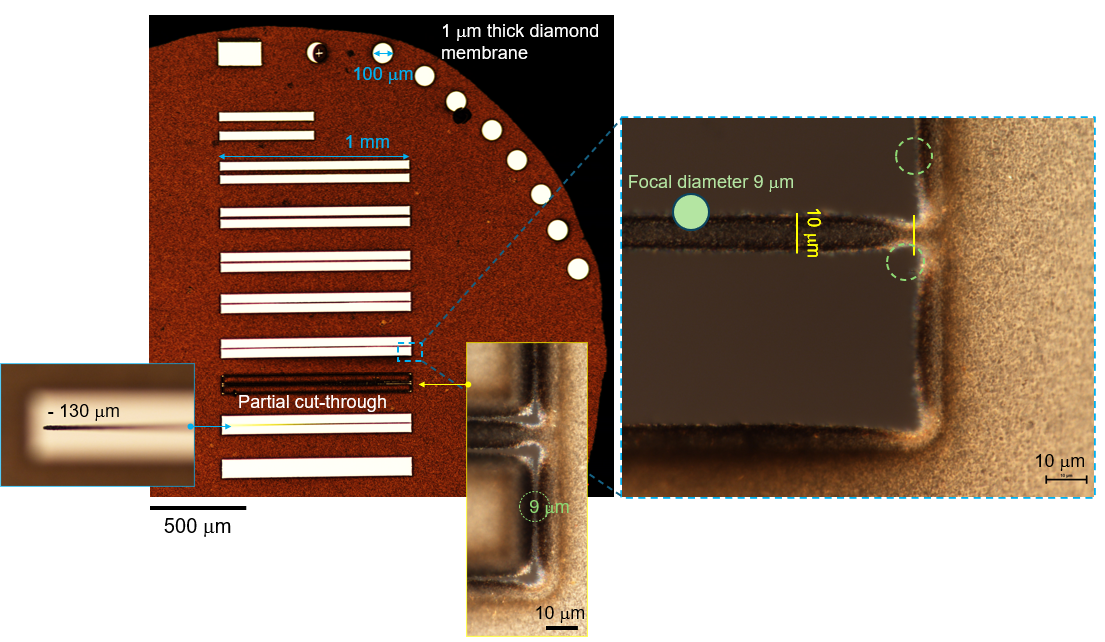}
\caption{\label{f-fs} Optical images of laser cut-out of 1-mm-long bridges in 1-$\mu$m-thick diamond membrane (free-standing over $\sim 3$~mm diameter hole). Laser conditions: $t_p = 230$~fs, $\lambda_l = 1030$~nm at 200~kHz and pulse density 300~mm$^{-1}$ or $\sim 3$ pulses per focal spot (Pharos in the WOP fs-fab station). Pulse energy $E_p = 300$~nJ (on sample), focused with $NA = 0.14$, focal diameter $2r = 1.22\lambda/NA = 9~\mu$m. Fluence per pulse $F_p = 0.474$~J/cm$^2$ and average irradiance $I_p = F_p/t_p = 2.06$~TW/cm$^2$. 
A full cut out was achieved in several passes passes (scan 10 times for extra redundancy).
 }
\end{figure*}

\subsection{Thin $\sim 1~\mu$m diamond membranes}

Example of laser ablation approach to define opto-mechanical structure with central 0.3-mm diameter disk suspended on two 1-mm-long arms of 10~$\mu$m width defined in 1-$\mu$m-thick membrane of diamond is shown in Fig.~\ref{f-cut}. Semi-circular $\sim 100~\mu$m wide incisions close to the membrane's attachment to Si substrate were made by fs-laser 230~fs/1030~nm ablation. The released internal stress formed a cylindrical arc of $\sim 80~\mu$m height over $2.5$~mm cross-section. The curvature of the cylindrical arc is $1/r \simeq 1/7.95$~mm$^{-1}$ (see bottom inset in Fig.~\ref{f-cut}). 

The protocol of laser cutting by ablation was developed on a diamond membrane without stress release by incisions. To define 1-mm-long bridge, one window was cut out by scanning at single fs-pulse fluence above the graphitisation threshold of 0.3~J/cm$^2$ and below the ablation threshold of 3~J/cm$^2$~\cite{ALI} (see, Fig.~\ref{f-fs}). The scanning speed, laser repetition rate and pulse energies were chosen to have at least two-three repeated cycles for a cut (Fig.~\ref{f-fs}). Once the first window was out, the second window is started with a shift determining the final beam width. Laser scan was started along the beam and progressed to the outside direction from the bridge. In this way, any membrane's movements due to stress release were minimised, which decreased the likelihood of unexpected lateral repositioning of the bridge in respect to the laser path, which would result in damaging the bridge/beam. The narrowest bridge of 10~$\mu$m was reliably fabricated. This protocol was used for the longer beam with a disc in the middle (Fig.~\ref{f-cut}). Carbonisation of the bridge and regions near the cut was apparent from color change. Internal stress of the diamond film was evident when 1-mm-long beam was separated at the one end from the rest of membrane. It moved $\sim 130~\mu$m from the plane of original membrane (Fig.~\ref{f-fs}). The final structures with longer span (Fig.~\ref{f-cut}), were made using more tight focusing with $NA = 0.45$ lens at higher fluence $\sim 3$~J/cm$^2$ in order to have more narrow width of graphitisation affected rim.        

The method of progressing from diamond graphitisation to oxidation (burning) is a debris-free method of 3D structuring and laser machining~\cite{23ol1379}. The energy deposition (dose) should be optimised for controlled removal of diamond into \ce{CO2}. The presented cutting of long micro-bridges was adopted to reduction of the width of the bridge by side-edge burning with precise control of lateral beam scanning. In this way, the most of the laser focal region is not irradiating the bridge/beam itself. This avoids direct linear momentum deposition (a push by $I/c$ photon pressure), which can break the beam. The Rayleigh length at the used focusing conditions $z_R = \frac{\pi r^2}{\lambda} = 247~\mu$m and allows to side-burn/polish micro-beams (waist/radius was $r = 9~\mu$m). With the used $NA = 0.14$ objective lens and $9~\mu$m focal spot, reliable 1-mm-long beams were made down to 8-10~$\mu$m. The major limitation is defined by the grain structure of CVD diamond membrane where sub-micrometer sized grains are not uniformly carbonized and burned (see Fig.~\ref{f-burn}). The grain structure of CVD diamond is also apparent in dry plasma etched regions near gratings (Fig.~\ref{f-x}). To remove graphitised carbon from edges of the laser cut of diamond membrane, \ce{H2} plasma can be used, which is more selective to $sp^2$ rather $sp^3$ bonding~\cite{etch}. The lower oxidation threshold of graphite vs. diamond in \ce{O2}~\cite{paci} could be utilised to reduce amount of graphitisation after laser cut. In fs-laser ablation of bulk high-pressure high-temperature (HPHT) diamonds at comparable high pulse fluence at 800~nm wavelength, graphitisation was found negligible~\cite{Zalloum}. The $sp^2$ regions in fs-laser inscribed modifications inside diamond are located along nano-cracks~\cite{Ashikk}.

\section{Discussion}\label{disco}

\subsection{An apparent anisotropy of $T$ (hence $A$) due to form-birefringence}

One clear evidence from 4-pol. spectra is its complex structure, where the dichroic ratio $R_d = \frac{A_\parallel}{A_\perp}$ is changing (Figs.~\ref{f-6780}, \ref{f-4120}) at the different parts of IR spectrum; here $A_{\parallel,\perp}$ is absorbance parallel or perpendicular to the selected direction. The dichroic ratio reveals orientation of the absorbing species in IR and widely used in IR chemical finger printing spectral region~\cite{IR}. The diamond membrane is isotropic and $R_d = 1$ is expected, especially at the regions were there are no specific absorption bands. However, a clear distinction between regions $R_d = \frac{A_\parallel}{A_\perp} = \frac{A_{90}}{A_{0}}\equiv \frac{A_o}{A_e}>1$ (at $90^\circ$ incidence azimuth of E-field polarization, $E$ is along the diamond ridges/planes of the grating, so, the o-beam or $A_{blue}>A_{black}$) as well as $R_d<1$ are observed at wavenumbers of high and low intrinsic absorbance parts of the spectrum, respectively (Fig.~\ref{f-4120}(a)). The absorbance dichroism revealed by 4-pol. analysis has origin in material structural organisation (a grating) which is made out of isotropic poly-crystalline absorbers. The form-birefringence, in theory, should not cause absorbance changes since it only affects state of polarization (ellipticity) of the outcoming beam but not energy losses/absorption. The light scattering, on the other hand, could contribute to the losses and affect the spectral lineshape of the absorbance bands~\cite{Hummel}. 

One possible contribution to enhanced absorbance in a grating structure is due to light intensity redistribution at the interface diamond-air between the walls. This is governed by the boundary conditions for the normal and tangential components of the displacement $D^{(n)}_1 = D^{(n)}_2$ and $E^{(t)}_1 = E^{(t)}_2$ at the interface for a non-conductive medium-1 (diamond) with surface charge density $\sigma = 0$ and air (medium-2). A change of dichroism from positive to negative occurs at the two sides in respect to the resonance peak~\cite{21as1544}. However, in non-absorbing spectral range the apparent dichroism stems from the form-birefringence.   

\begin{figure*}[tb]
\centering\includegraphics[width=0.95\textwidth]{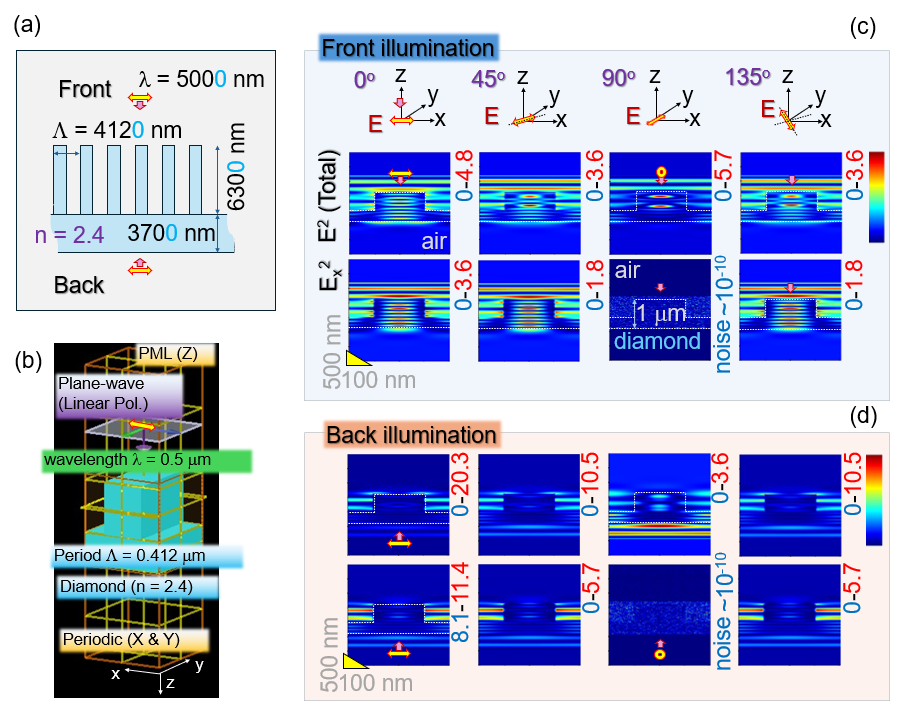}
\caption{\label{f-toy} (a) FDTD numerical model (not to scale) of the diamond grating (duty cycle 0.5) on a membrane; diamond refractive index $n=2.4$. The Maxwell’s scaling is applied: reduce dimensions by factor of $10^\times$, membrane from 1~$\mu$m (actual) to 1~$\mu$m (the last digit in \blue{``0''}, is dropped out for the size definition) as well as for the wavelengths of source from $\lambda = 5~\mu$m (2000~cm$^{-1}$) to 500~nm. (b) FDTD model for optical domain (not IR) with perfectly matching layer (PML) conditions for z-axis and periodic for x- and y-axis was used. (c) Total intensity $E^2$ and $E_x^2$ (the normal (n) field in the plane of mapping) for four azimuths of incident linear polarization; incident intensity $E_0^2 = 1$. Plane wave illumination from front (top of grating). (d) Same as (c) only for the back-side illumination.     }
\end{figure*}

The dichroic function $f_D$ measured experimentally is linked to the $\gamma$ angle between the electric dipole-transition moment $M$ and selected direction of $E$-field orientation: $f_D = \frac{R_d-1}{R_d+2}=\frac{1}{2}(3\cos^2\gamma-1)$; with the absorbance maximum $A_\parallel$ when $\gamma = 0^\circ$ and zero when $\gamma = 90^\circ$ at $A_\perp$~\cite{IR}. The $f_D = 0$ when $R_d = 1$ and there is no anisotropy in absorption. At this condition, $\left\langle\cos^2\gamma\right\rangle = 1/3$ or $\gamma\simeq 54.74^\circ$. The $f_D$ is plotted for different period gratings in Fig.~\ref{f-Rd}. For the smallest period of 820~nm, the $f_D$ is oscillating with a structure of $FSR$. However, a change from $f_D<0$ to $f_D>0$ is clearly discernible for the 4.12~$\mu$m grating between positive and negative dichroism. At the intrinsic absorbance region of diamond $f_D\simeq 0.5$, i.e., $\cos\gamma = \sqrt{2/3}$ or $\gamma\approx 36^\circ$ (positive linear dichroism). There were no any specific $f_D$ changes near the highly reflective Reststrahlen spectral region at the degenerate TO and LO phonon bands, which for diamond are $\omega_{TO}\approx\omega_{LO}$ at $\sim 1332$~cm$^{-1}$~\cite{THON}, which is also a characteristic Raman line of diamond~\cite{1332}.

\subsection{Numerical model: optical intensity distribution}

In order to model light intensity distribution relevant to a comparatively large grating on a diamond membrane, which has dimensions and structures $10-20\lambda$ in size, where $\lambda$ is the wavelength, 
a 10$^\times$ down-scaling was applied in order to apply FDTD (see Fig.~\ref{f-toy}(a)). FDTD model for the visible wavelength 500~nm was setup up with periodic boundary conditions in the lateral dimensions and the perfectly matching layer (PML) along light propagation to remove scattering and reflections Fig.~\ref{f-toy}(b). Figure~\ref{f-toy}(c) shows cross sectional view of the light intensity enhancement in the form-birefringent structure for top-illumination. The $E_x^2$ corresponds to the normal $(n)$ component and is shown separately from the overall intensity. The back-illuminated intensity cross-section is presented in Fig.~\ref{f-toy}(d) and the local near-field distribution is apparently different. As required by boundary conditions, the $(n)$ component has largest enhancement at the interface diamond-air at the side-walls of the grating structure. Also, there is enhancement of intensity at the exit interface for propagation from diamond-to-air when there is no phase change for the reflected light. This illustrates the difference of volume where high intensity is localised, hence, were a strong absorption for the $R$ and $T$-modes could be expected. The form-birefringent structure is responsible for this energy redistribution and 
influence transmission (hence absorption) due to different proportions of energy distributed between the diamond and air. The modeling of plane wave irradiation and periodic boundary conditions present the ideal case when all scattered/diffracted light is accounted for. In real experiments, the illumination/collection $NA = n\sin\alpha$ of the objective lens plays a role since only light within a cone $2\alpha$ will be measured/collected; here $n$ is the refractive index of air (for dry objective lens).

\section{Conclusions}

The $T$ and $R$ spectral of diamond micro-gratings at IR spectral range showed polarization sensitivity at the intrinsic absorption band of diamond where change of dichroic ratio was changing sign (positive linear dichroism was at the intrinsic absorption region). In a 4-pol. measurement, this corresponded to a reversal between the high and low absorbance. Absorbance was also accessed from the reflectance spectra, and confirmed dichroism reversal. This can be useful for structures when transmittance is not possible to measure directly, which is very common case in complex and integrated micro-structures.       

For laser cutting of micrometers-thick diamond membranes, the approach of graphitisation (0.3~J/cm$^2$ threshold) and oxidation is adopted to reduce threshold of ablation (3~J/cm$^2$) and was applied for fabrication of micro-bridges of $\sim$1~mm long, $10~\mu$m-wide in $1~\mu$m-thick diamond membrane. Stress release cuts were essential to define such micro-bridges. 

\small\begin{acknowledgments}
S.J. acknowledges support via ARC DP240103231 grant and Nanophorb (2018-2022) - Nanofabrication technologies towards advanced control of the photon angular momentum - from the National Center for Scientific Research (Le Centre National de la Recherche Scientifique), France. Design of micro-bridges for opto-mechanical applications was provided by Etienne Brasselet. H-H.H. and S.J. are grateful for a research stay at the Laser Research Center, Vilnius University, in 2025. J.M. acknowledges support via JST CREST (Grant No. JPMJCR19I3) and KAKENHI (No.22H02137). M.R. was supported via KAKENHI (Grant No. 22K14200) and SAKIGAKE (Grant No. JPMJPR250E). Part of the supplementary information was acquired using synchrotron-based FTIR microspectroscopic technique on the Infrared Microspectroscopy (IRM) beamline at the Australian Synchrotron, part of ANSTO.  
\end{acknowledgments}

\bibliography{aipsamp}

\appendix
\setcounter{figure}{0}\setcounter{equation}{0}
\setcounter{section}{0}\setcounter{equation}{0}
\makeatletter 
\renewcommand{\thefigure}{A\arabic{figure}}
\renewcommand{\theequation}{A\arabic{equation}}
\renewcommand{\thesection}{A\arabic{section}}

\section{Anisotropy in $T$ and $R$-modes: no fringe removal}

\begin{figure*}[h!]
\centering\includegraphics[width=.7\textwidth]{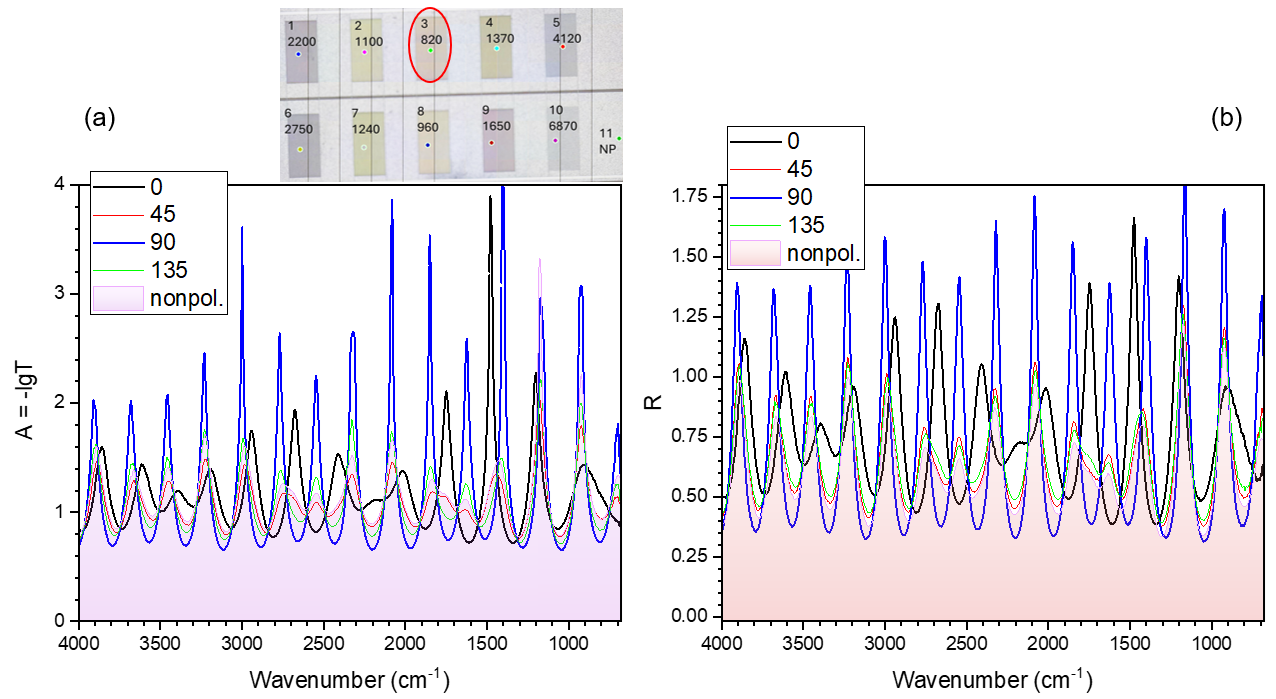}
\caption{\label{f-820} Transmission mode (a) and reflection (b) spectra at 4-polarizations (set before sample) for the $\Lambda = 0.82~\mu$m grating. Free spectral range $FSR = 236$~cm$^{-1}$ (from (b)). 
}
\end{figure*}

Figure~\ref{f-820} presents $T,R$-spectra for 820~nm period grating (0.5 duty cycle). This is sub-wavelength structure showing strong oscillatory signals due to thickness of entire structure on the membrane being only 4-10 times larger than the wavelength of the light source.

\begin{figure*}[h!]
\centering\includegraphics[width=.65\textwidth]{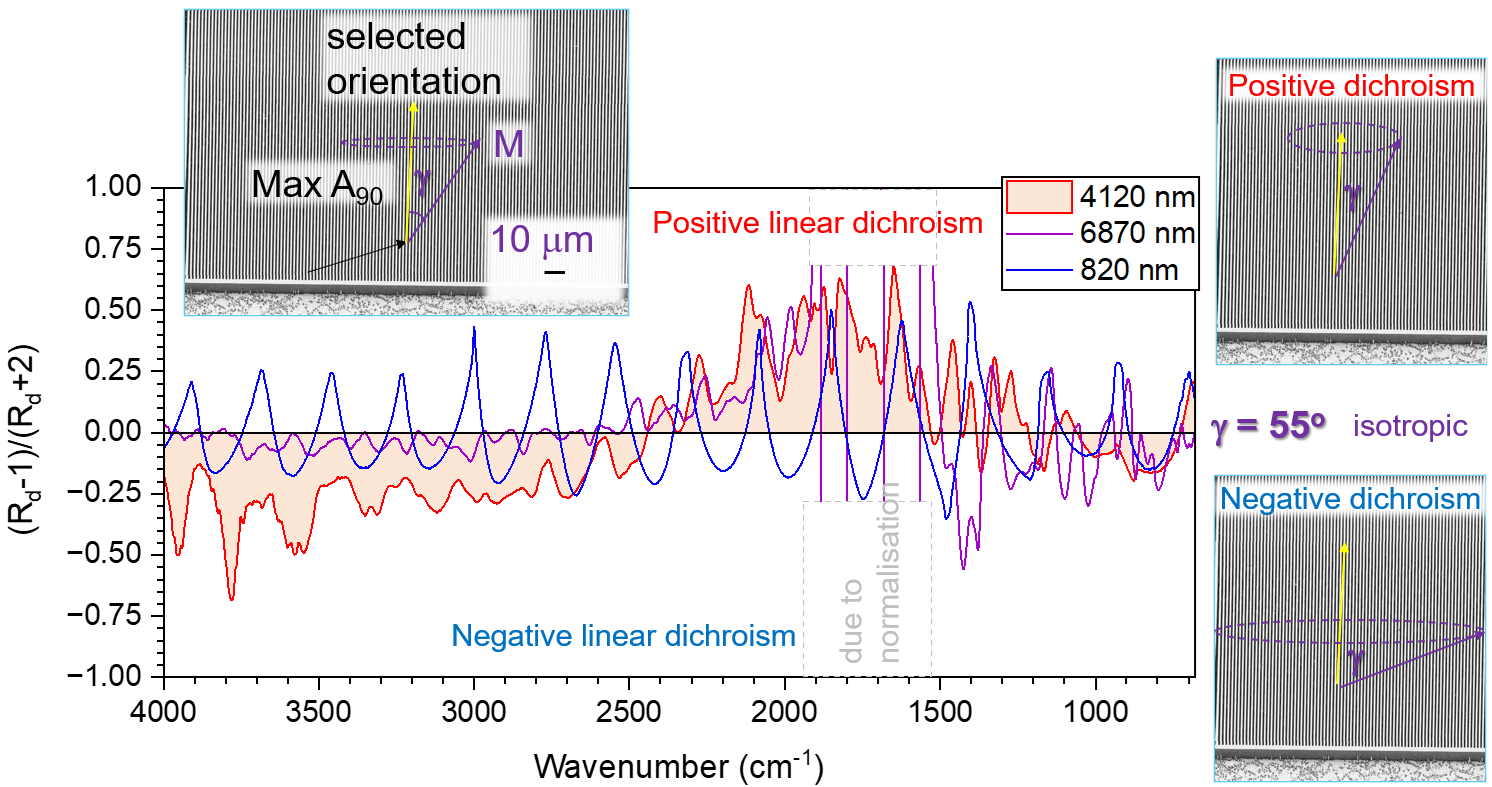}
\caption{\label{f-Rd} The dichroic function $f_D = \frac{R_d-1}{R_d+2}$ for three grating periods $\Lambda$. The normalisation of absorbance $A$ caused artificially oscillating large signal for the $\Lambda = 6.87~\mu$m grating (marked). The inset shows SEM image of diamond grating with the selected orientation along the ridges (maximum of absorption designated as $A_{90}$) and the transition dipole momentum $M$; $\gamma$ is the angle between them.
}
\end{figure*}

Figure~\ref{f-Rd} shows spectra of dichroic function with change of linear dichroism at the intrinsic absorption window of diamond. 

\begin{figure*}[h!]
\centering\includegraphics[width=.55\textwidth]{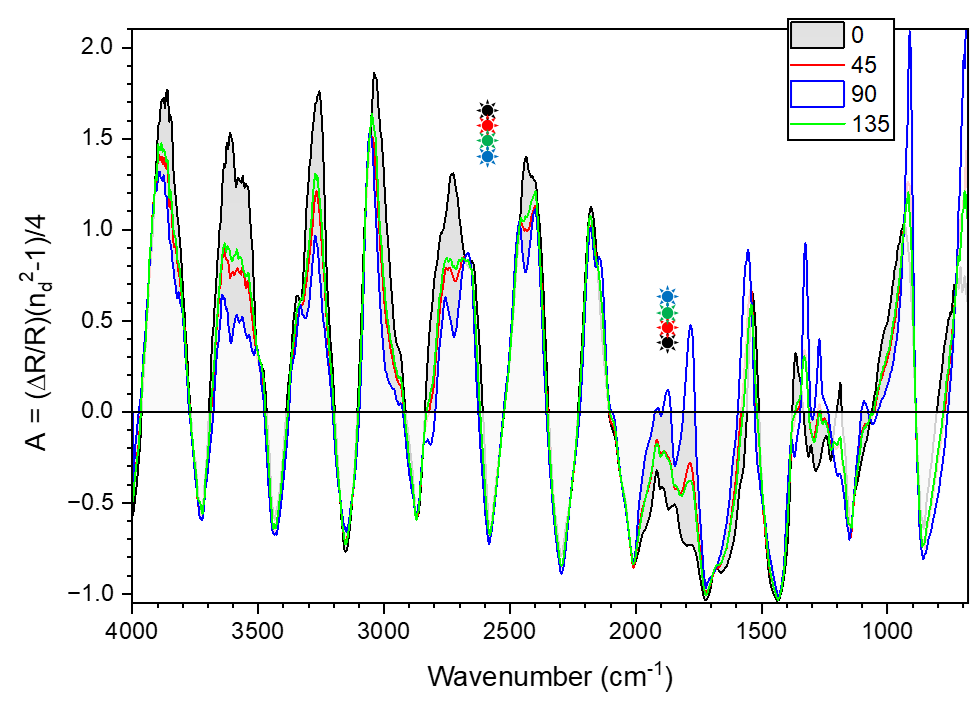}
\caption{\label{f-ref} Absorbance calculated from experimentally determined differential reflectance $A = \frac{\Delta R}{R_D}\times\frac{n_d^2-1}{4}$ for 4.12~$\mu$m period grating at four linear polarizations (at incidence onto sample); $n_d = 2.35$ and the spectra on the membrane (plasma thinned region) was used as the reflectance reference $R_D$. Reversal of dichroic ratio sign is recognisable at the diamond absorption band (compare with Fig.~\ref{f-Rd}). Negative $A$ region is caused by stronger reflectance from diamond than gold (used for normalisation). Grey-shading is used between $A_{90}$ and $A_0$ as eye-guide to the change of dichroic ratio. 
}
\end{figure*}

Figure~\ref{f-ref} presents absorbance $A$ spectra calculated from the differential reflectance $\Delta R/R_D$, which represents absorbance only in the grating structure since for the reference a thinned diamond membrane region was used (reflectance $R_D$). 

\section{Kramers-Kronig determination of refractive index from reflectance spectra}

Assessing absorbance from reflectance $R$ can also be achieved by Kramers-Kronig analysis. The complex amplitude of the reflected light wave $r^*$ is a function of the phase difference between the incident and reflected waves $r^* = |r|e^{i\phi}\equiv\sqrt{R}(\cos\phi +i\sin\phi)=\frac{n-i\kappa-1}{n-i\kappa+1}$, where $\phi$ is the phase angle. The reflectance $R=|r|^2 = \frac{(n-1)^2+\kappa^2}{(n+1)^2+\kappa^2}$. The refractive index $n+i\kappa$ is given by~\cite{Okamura}:
\begin{equation}
n = \frac{1-R}{1+R-2\sqrt{R}\cos\phi},~~~~ \kappa = \frac{-2\sqrt{R}\sin\phi}{1+R-2\sqrt{R}\cos\phi}.    
\end{equation}

For experimentally measured spectra $R(\omega)$, the phase $\phi$ is obtained for each frequency value $\omega_i$ from the Kramers-Kronig relation~\cite{Okamura}:
\begin{equation}\label{e-kk}
\phi(\omega_i) = \frac{2\omega_i}{\pi}\int_0^\infty\frac{\ln\sqrt{R(\omega)}-\ln\sqrt{R(\omega_i)} }{\omega_i^2 - \omega^2} d\omega.   
\end{equation}
This integral has poles at $\omega_i = \omega$ and can be calculated using the double Fourier transform method~\cite{Shimadzu}:
\begin{equation}
 \phi(\omega_i) = 4\int_0^\infty\cos(\omega_it)dt\times\int_0^\infty\ln(\sqrt{R(\nu)}\sin(2\pi\nu t)d\nu,
\end{equation}
where $\omega=2\pi\nu$ is the cyclic frequency. By measuring wide IR spectrum range benefits recovery of $n,\kappa$ and allow to estimate absorbance or optical density (from reflectance only) $A=OD = d\times \frac{4\pi\kappa}{\lambda}$, where $d$ is the thickness and the absorption coefficient $\alpha = \frac{4\pi\kappa}{\lambda}$~[cm$^{-1}$].

Another computable approach to estimate Eqn.~\ref{e-kk} is following~\cite{KK,Stern,Kozak}:
\begin{equation}
    \phi(\omega_i) = -\frac{1}{2\pi}\int_0^\infty\ln{\left|\frac{\omega-\omega_i}{\omega+\omega_i} \right|}\times \frac{d}{d\omega}[\ln\sqrt{R(\omega)}-\ln\sqrt{R(\omega_i)}] d\omega. 
\end{equation}

\section{Graphitisation of diamond}

\begin{figure*}[h!]
\centering\includegraphics[width=.7\textwidth]{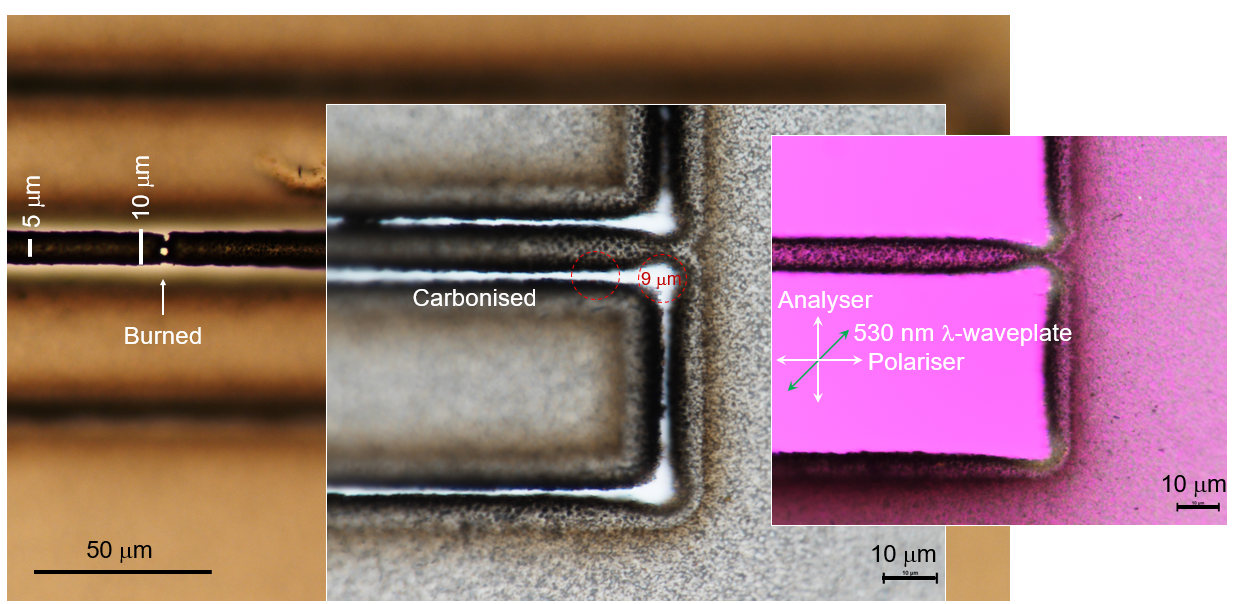}
\caption{\label{f-burn} Optical images (actual color) of beam/bridge fs-laser defined by carbonisation of diamond and subsequent oxidation of 1-$\mu$m-thick diamond membrane. The cross-polarized image with $\lambda = 530$~nm shows no stress build in near the laser cuts. Laser-fab conditions are as for structures in Fig.~\ref{f-fs}.
}
\end{figure*}

Figure~\ref{f-burn} shows close up optical images of bridges fs-laser cut out from diamond membrane shown in Fig.~\ref{f-fs}. Cross polarized imaging showed that there was no stress induced at the vicinity of anchoring sites of the beam/bridge. Whiskers of micrometer size were recognisable along the laser oxidised tracks and could be due to the diamond orientation controlled  propensity to oxidation~\cite{Quar}. Degree of graphitization can be determined using Raman and IR spectroscopes which are sensitive to changes in $sp^2$ and $sp^3$ bonding and corresponding C-C vibration bands~\cite{SHAR}.  

\begin{figure*}[h!]
\centering\includegraphics[width=1\textwidth]{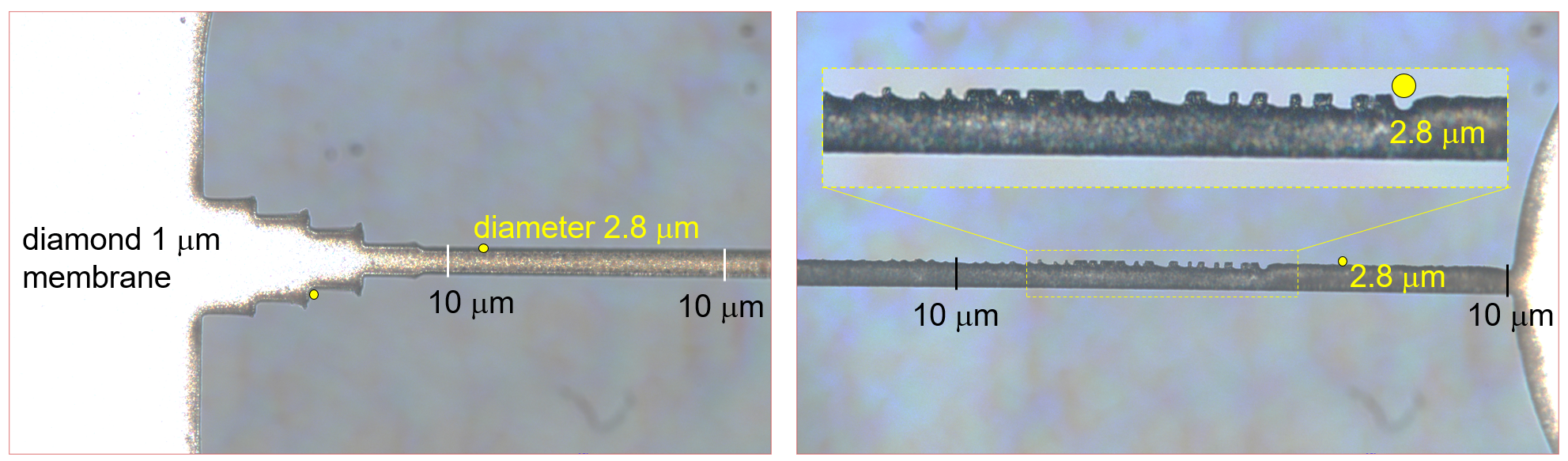}
\caption{\label{f-edge} The edge trimming at higher laser power (but on the same beam trajectory) after the initial cutting of an entire structure creates defects on a smooth edge when micro-beam/bridge structure moves laterally during the laser scan.
}
\end{figure*}

At more tight focusing (Fig.~\ref{f-edge}) the final cut out of a suspended disk on two side beams of 1~mm length was made. Attempt was made to thin the width of 10~$\mu$m beam/bridge by extra passes of the laser beam at higher pulse energy (power). At higher laser power and tighter focusing, the lateral force can push out high refractive index material (diamond $n\simeq 2.4$) as well as pull into laser beam depending of the actual position of object and focal spot~\cite{07oe13310}. 

\begin{figure*}[tb]
\centering\includegraphics[width=0.9\textwidth]{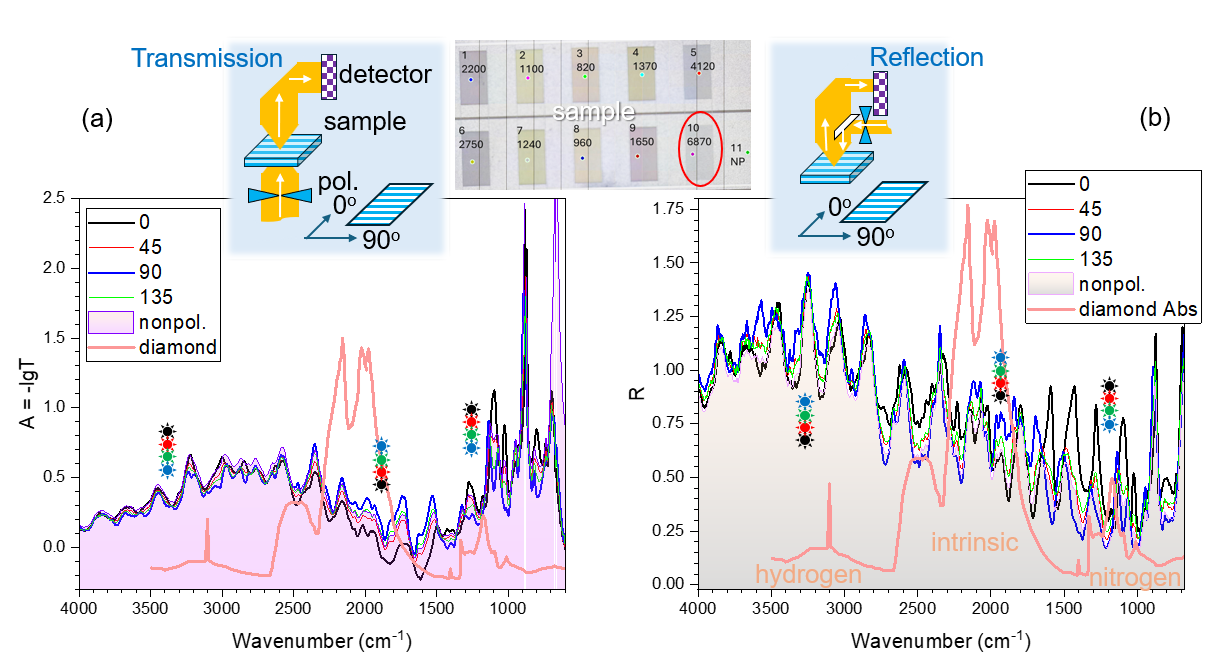}
\caption{\label{f-6780} Transmission $T$ (absorbance $A = -\lg T$ in (a)) and Reflection (b) modes of the FTIR spectra at four different linear incident polarizations with azimuth $\theta= 0^\circ,45^\circ,90^\circ, 135^\circ$ measured from the same area $100\times 100~\mu$m$^2$ defined by an aperture; $\theta = 0^\circ$ is along the vertical (y-axis) and along the extraordinary e-ray direction with refractive index $n_e$ ($n_e < n_o$ in the form-birefringent structure of grating). The star-markers show high-low signal alignment.
FT-IR Bruker Hyperion 3000 with an $15^\times$ magnification Cassegrainian objective lens with numerical aperture $NA = 0.4$ 
(cone angle $\sim 24^\circ$). Incident IR light from Globar-source is polarized in front of sample before incidence and not polarization discriminated at the detector. Inset shows samples; the measured grating has period $\Lambda = 6.87~\mu$m (1456~cm$^{-1}$). The typical absorption spectral profile of a diamond (Bruker~\cite{Bruk}) is shown on top of 4-pol. spectra. Transmitted signal was normalised to transmission without sample (air) and reflected signal was normalised to reflection from gold.}
\end{figure*}

\begin{figure*}[h!]
\centering\includegraphics[width=0.9\textwidth]{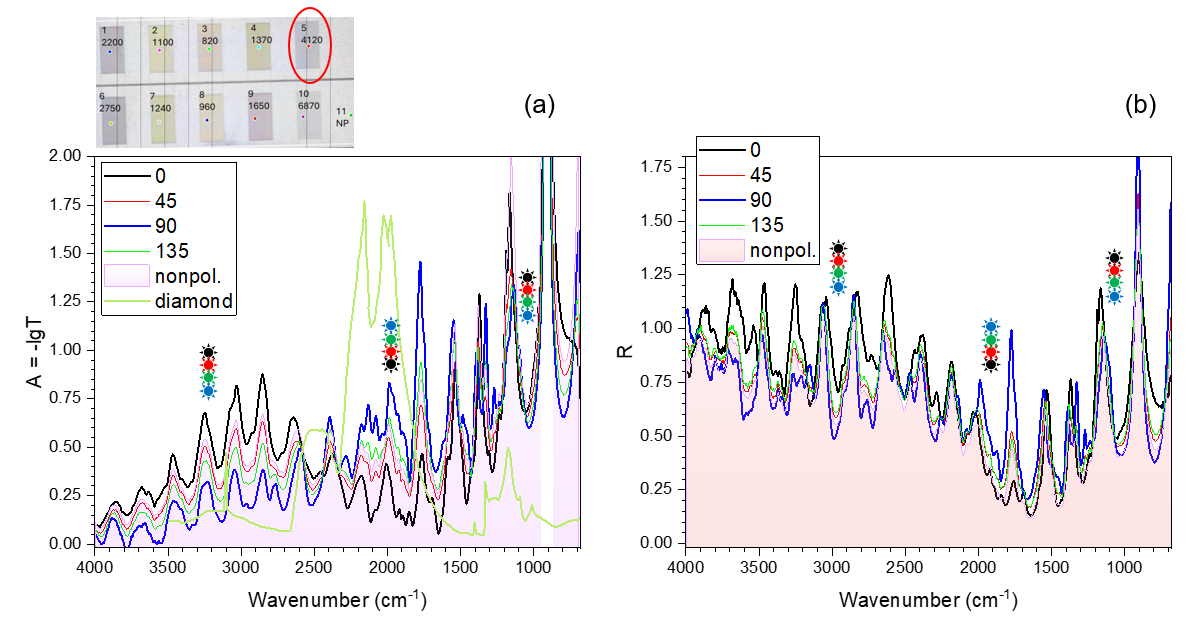}
\caption{\label{f-4120} Transmission mode (a) and reflection (b) spectra at 4-polarizations (before sample) for the $\Lambda = 4.12~\mu$m (2427~cm$^{-1}$) grating. Absorbance spectral profile of a typical diamond is overlayed on absorbance $A = -\lg T$ in (a). The star-markers show high-low signal alignment. }
\end{figure*}


\end{document}